# Characterization and Manipulation of Mixed Phase Nanodomains in Highly-strained BiFeO$_3$ Thin Films

*Lu You,[†] Zuhuang Chen,[†] Xi Zou,[†] Hui Ding,[†] Weigang Chen,[†] Lang Chen,[†] Guoliang Yuan,[‡]\* and Junling Wang[†]\**

[†] School of Materials Science and Engineering, Nanyang Technological University, Singapore, 639798

[‡] Department of Materials Science and Engineering, Nanjing University of Science and Technology, Nanjing, 210094, China

\*E-mail: yuanguoliang@mail.njust.edu.cn, jlwang@ntu.edu.sg

ABSTRACT

The novel strain-driven morphotropic phase boundary (MPB) in highly-strained BiFeO$_3$ thin film is featured by ordered mixed phase nanodomains (MPNs). Through scanning probe microscopy and synchrotron X-ray diffraction, eight structural variants of the MPNs are identified. Detailed polarization configurations within the MPNs are resolved using angular-dependent piezoelectric force microscopy. Guided by the obtained results, deterministic manipulation of the MPNs has been demonstrated by controlling the motion of the local probe. These findings are important for in-depth understanding of the ultrahigh electromechanical response arising from phase transformation between competing phases, enabling future explorations on the electronic structure, magnetoelectricity and other functionalities in this new MPB system.

The recent discovery of a strain-driven morphotropic phase boundary (MPB) has spurred resurgent interest in multiferroic BiFeO$_3$ (BFO).[1] A novel tetragonal-like ("T"-like) phase with giant axial ratio can be stabilized under large compressive strain when grown on LaAlO$_3$ (LAO) substrate.[2] With increasing film thickness, another rhombohedral-like ("R"-like) phase starts to emerge to release the elastic energy. It mixes with the parent "T"-like phase to form stripe-like patches dispersed in the "T"-like matrix, which is termed mixed phase nanodomains (MPNs). The relatively small difference in the energy scale between these phases makes them very susceptible to external stimulus, e.g. electric field or stress, leading to ultrahigh electromechanical response comparable to ferroelectric relaxors.[3] More recently, ferromagnetism was also observed at the competing phase boundaries, which can be written or erased using electric field.[4] These exciting functional properties may lead to applications in low-power sensors, actuators and memories.

Previous detailed structural studies have revealed that the "T"-like matrix phase is in fact of monoclinic $M_C$ symmetry with the polarization lies in the {100} planes,[5,6] in contrast to the conventional "R"-like BFO (e.g. monoclinic $M_A$ phase under small compressive strain, monoclinic $M_B$ phase under small tensile strain as well as rhombohedral phase in bulk), whose polarization vectors are confined within the {110} planes.[7] On the other hand, the "T"-like and "R"-like phases in the MPNs are found to be highly distorted and greatly different from their parent counterparts, probably with lower symmetry (we refer them as $T_{tilted}$ and $R_{tilted}$ phases in the following context).[8,9] The emergence of low symmetry phases with competing energy scale is the common signature of MPB systems. However, in conventional MPBs induced by chemical alloying, characterizing the local domain structures of MPNs is rather challenging due to the compositional inhomogeneity. In comparison, the strain-driven MPB in BFO has distinct MPNs with ordered alternating "T"/"R"/"T"/"R" phase lamellae, which provide a perfect platform for in-depth study of the ferroelectric orders at nanoscale. However, due the complexity of the structural variants of the MPNs, a clear picture of the local domain structures remains elusive. In this report, we presented the first detailed characterization of the MPNs in highly-strained BFO thin films by a combination of scanning probe microscopy and synchrotron X-ray reciprocal space mapping (RSM). The local polarization configurations in the MPNs were mapped out using angle-resolved piezoelectric force microscopy (AR-PFM).[10-11] Clarifying the polarization orientations at the boundaries between these two phases helps us to gain deeper insight into the polarization rotation path among low-symmetry phases in BFO during the phase transition process, which is directly related to how the large mechanical response is generated and how we can control it. As an example, we also demonstrated deterministic writing and erasing of the MPNs by controlling the interaction between the local polarization directions and the electric field produced by the biased PFM probe. These results should facilitate further explorations on the emergent functional properties of the multiferroic MPB system with electrical tunability.

**Identification of Different MPN Variants.** Typical topographic images from two different locations of a 60 nm-thick BFO film grown on LAO substrate were shown in Figure 1a&b, together with the corresponding in-plane (IP) PFM images (Fig. 1c&d). The smooth area is featured by unit-cell-height terraces, comprised of the $M_C$ phase. Stripe-like MPNs with various orientations (denoted by color boxes) seemingly randomly distributed within the matrix of the $M_C$ phase. By scrutinizing a large film area, it can be concluded that the orientation of the MPN arrays are primarily along eight possible directions in the film plane. Thus, we can define different types of MPNs by vectors perpendicular to the stripe patterns, as denoted by the arrows.[12] These structural vectors tilt away from the IP <100> axes by about 10°. To confirm this, cross-sectional RSM was carried out by fixing the L at the out-of-plane (OP) lattice constant of the $R_{tilted}$ phase in the MPNs (Supporting Information, Figure S1). As shown in Figure 1g, the eight peaks in the H-K map unambiguously indicate eightfold degeneracy of the structural variants of the MPNs, consistent with the structural vectors drawn from the topographic images. The difference between the variants with positive and negative superscript can be seen from the local height profiles of the MPNs. Taking MPN$_1^+$ and MPN$_1^-$ as an example (red boxes in Fig. 1a&b), both of them display a scrubboard-like morphology, as the [001] axes of the T$_{tilted}$ and R$_{tilted}$ phases rotate away from the film normal by ~1.5° and ~2.7°, respectively (Fig. 1e).[4, 8-9] However, the tilting directions are opposite due to the different arrangements of the nanodomains as indicated in Figure 1f.

**Resolving Polarization Configurations in MPNs.** After establishing the eight possible variants, we can now look closely into the ferroelectric domain structures in the MPNs. In the IP PFM images, the M$_C$ phase matrix exhibits well-aligned stripe domains with the domain wall lying along the <110> axes, in agreement with previous reports.[5] In contrast, the domain patterns in the MPNs more or less coincide with the surface morphologies, with both the T$_{tilted}$ and R$_{tilted}$ subdomains showing similar contrast in each MPN. However, for different MPNs, varying contrasts suggest different polarization orientations. Generally, the color tones in the MPNs are consistent with the surrounding M$_C$ matrix phase. For example, in Figure 1c, both display a yellow tone, while Figure 1d shows an overall purple tone. This observation indicates that the emergence of the MPNs takes place during cooling process from high temperature.[13,14] In order to minimize the elastic energy, the polarizations in the MPNs should adopt proximate rotation path with regard to the matrix phases. Therefore, the polarization direction in the tilted phases should be close to the M$_C$ matrix and show similar contrast in PFM images.

To determine the exact polarization configurations in the MPNs, AR-PFM was performed by recording the IP PFM images at different azimuthal angles between the probe cantilever and the [100] axis of the sample. As we know, in a PFM image, the amplitude signal is proportional to the magnitude of the polarization component along the cantilever normal, whereas the phase signal reflects the sign of this polarization component. It is thus possible for us to identify the IP orientation of the polarization vector by finding out at which rotation angle the phase signal changes sign and the amplitude signal goes to minimum. At this angle, the polarization vector will be just parallel to the cantilever axis. The results are illustrated in Figure 2. Here we focused on two specific MPN variants, namely MPN$_2^-$ and MPN$_3^-$, as indicated by the arrows of the structural vectors. The cantilever was fixed along [100] direction, while the sample was rotated clockwise from 0° to 90°. As shown in Figure 2a, for MPN$_3^-$ the contrast in IP PFM images didn't change within the entire rotation range, maintaining a yellow tone. However, the color tone of MPN$_2^-$ changed from yellow to purple at around 45°. According to the OP PFM images, the polarization of the entire film was uniformly pointing downwards (Supporting Information, Figure S2). By decoupling the IP PFM image into phase and amplitude signals, their angular dependence can be plotted for both T$_{tilted}$ and R$_{tilted}$ phases in each MPN, as shown in Figure 2b&c. Clearly, the phase ($sin\theta$) signals in MPN$_3^-$ remained constant at 1, whereas those in MPN$_2^-$ changed from 1 to -1 during sample rotation. By fitting the data, it was found that the phase reversal occurs at 34°±3° and 45°±3° for T$_{tilted}$ and R$_{tilted}$ phases in MPN$_2^-$, respectively. Thus, the exact orientations of the polarization vectors in MPN$_2^-$ can be determined as indicated by the arrows in the coordinate of Figure 2a. As for the amplitude signal, since it is highly sensitive to the tip-sample contact and the wear-out of the tip, it cannot be quantitatively compared among separate scans.[10] However, within each scan, the amplitude signal is relatively consistent. Therefore, it can be normalized with respect to the signal of the M$_C$ phase whose polarization orientations are already known. The results deduced from the normalized amplitude signals are in good agreement with the phase data, as shown in Figure 2c. It should be noted that the IP polarization of the R$_{tilted}$ phase in MPN$_2^-$ actually lies close to the [110] axis, raising a reminiscence of conventional "R"-like BFO. On the other hand, the IP polarization of T$_{tilted}$ phase lies at an intermediate angle between the polarization vectors of M$_C$ phase and the R$_{tilted}$ phase. At the same time, the M$_C$ phase surrounding MPD$_2^-$ consists of the two-variant stripe domains with the polarizations pointing along [100] and [010] directions. These two polarization variants form the quadrant within which the polarizations of the mixed phases are contained. This further proves that the mixed phases develop from the parent M$_C$ phase through proximate polarization rotation when cooling from high temperature. We also repeated the AR-PFM measurements on other samples. The results are consistent with what is reported here (Supporting Information, Figure S3). Using symmetry operation, the polarization vectors of the rest of MPNs can be deduced as illustrated in

Figure 3f. For simplification, only half of the MPN variants are shown with an average polarization vector lying ~40° away from the <100> axes for each MPN.

**Deterministic Control of MPNs by PFM Probe.** An immediate application of resolving the polarization directions in the MPNs is to guide us in controlling the polarization switching path and the MPN formation using electric field produced by the PFM probe. Such a technique has been previously shown to deterministically control the ferroelastic switching in "R"-like BFO films.[15] However, the situation is more complicated here as it involves not only the ferroelastic switching but also the electric field induced phase transition between the parent $M_C$ phase and the MPNs. Recent work by Vasudevan et al has studied the phase control of MPNs using PFM probe.[16] However, due to the absence of a detailed picture of the local polarization configurations, they were not able to correlate the local electric field with specific MPN variants. Thus, the control is not completely deterministic. Building upon the established polarization map, we will show below precise control of MPN variants by correlating the local field with the polarization direction. The underpinning mechanism can be understood in terms of the dynamic interaction between electric field and polarization during the phase transformation process.

A 30 nm-thick BFO film on LAO with mostly $M_C$ phase and negligible MPNs was selected for the demonstration (Figure 3a). It should be pointed out that even though the bottom electrode is absent, due to the large stray field generated by the probe, the transformation between $M_C$ phase and the MPNs can still occur without switching OP polarization.[17] As a result, only the IP component of the electric field needs to be considered for the interaction with the IP polarizations. Because the OP polarization of all the films are pointing downward, a negative bias (-9 V) was applied to the probe to compress the $M_C$ phase into MPNs. Following the notation in reference [15], the local IP electric field can be decomposed into components along fast and slow scan directions, as schematically shown in Figure 3a. Due to the movement of the probe, every scanned area will experience changes in field direction when the probe is passing through. Therefore, every new domain formed at the front of the probe will be constantly replaced by the other after the probe passes over it. The final state of the polarization direction will be determined by the electric field at the rear side of the slow scan direction. Since the fast and slow scan direction can be easily modified by changing the scan angle, the field-induced formation of MPNs thus can be selectively controlled. For example, when the scan angle is 0°, the slow scan direction points toward [0-10] direction, so is the final electric field. As a result, we should expect the formation of MPN with most polarization component lying along this direction. Surprisingly, this process resulted in $MPN_4^-$, which has a polarization component opposite to the final electric field direction as demonstrated in Figure 3b. Besides, based on the polarization map in Figure 3f, when the field is along [100] or [010] axis, two MPN variants ($MPN_4^-$ and $MPN_1^-$ in 0° scan angle case) are energetically degenerated with the same polarization component along the field direction. However, practically only one MPN is generated for each scan. This could be due to the asymmetric field distribution induced by the wear-out of the probe or the slight misalignment of the sample, which lifts the degeneracy. Afterwards, the MPNs were erased using a positive bias (+9 V), recovering the original $M_C$ phase. Again, we wrote the MPNs at the same location by rotating the scan direction for 30°, $MPN_1^-$ emerged instead of $MPN_4^-$ because now it has larger polarization component along the field direction. If the probe scans at 180°, the polarization direction of the induced MPN also reversed (Figure 3c). MPNs with larger polarization components along [100] axis can be obtained by scanning at 90° as shown in Figure 3d. Likewise, the rest of MPN variants can be produced by controlling the motion of the biased probe. However, all the polarizations of the field-induced MPNs are indeed opposite to the electric field directions at the rear side of the slow scan direction. This is completely different from previous study on field-directed ferroelastic switching in "R"-like BFO films with stripe domains,[15] but qualitatively agrees with the switching characteristics of BFO films with "bubble-like" domains.[18]

To better understand the polarization switching scenario during electric field directed phase transition, IP PFM images were recorded before and after the electrical writing process, as shown in Figure 4. The experimental conditions were the same as previous, and the scan angle here is 180°, that is, scanning upwards. However, $MPN_1^+$ was formed instead of $MPN_4^+$ shown previously (Figure 4c). This further supports the argument that the degeneracy between these two MPNs is broken due to the reason described above, leading to only one preferred MPN. According to our polarization map derived from AR-PFM study, the polarizations of the mixed phases in $MPN_1^+$ have a downward-pointing component, again, opposite to the electric field. This is verified in the IP PFM images. The pristine domain structure of parent $M_C$ phase is characterized by yellow/brown stripe pairs with the net polarization pointing upwards (Figure 4b). After electrical writing, the switched area shows a dominant purple contrast, indicating the net polarization has been switched to downward direction (Figure 4d). However, if we focus on the first few scan lines, it can be observed that the IP polarization remains the same as the pristine state, and the mixed phases formed in this region is $MPN_1^-$ instead. The only difference between the first few scan lines and the rest is that they didn't experience the electric field at the front side of the slow scan direction. Therefore, we believe that the polarization directions in the MPNs are determined by the electric field at the front side of the slow scan direction rather than that at the rear side. A two-step phase transition scenario during the electrical writing process thus can be envisioned, as illustrated in Figure 4e. Firstly, the domains of parent $M_C$ phase will experience electric field at the front side of the slow scan direction before the probe actually passes through them. This IP electric field should be large enough to align the polarization of the $M_C$ phase through ferroelastic switching, similarly to the "R"-like BFO case (Supporting Information, Figure S4). At this stage, the phase transition from $M_C$ phase to mixed phases will not take place because the driving force, namely, the OP electric field is negligible. When the probe passes through, the highly concentrated OP electric field will depress the polarization of the domain right beneath the probe, leading to the transformation of $M_C$ phase into a mixture of $T_{tilted}$ and $R_{tilted}$ phases. At this stage, the type of induced MPNs is determined by the polarization variants of the parent $M_C$ phase formed at the previous stage. As mentioned above, the polarization rotations in the mixed phases will remain proximate to the parent phase, without changing the contrast shown in the PFM images. Finally, after the probe passes the scanned area, the electric field at the rear side won't be able to switch the polarization anymore due to the elastic constraint imposed by the formed MPNs. As a consequence, the polarizations in the scanned domains are determined by the electric field at front side of the slow scan direction, which explains all the experimental observations.

**Conclusion.** The work presented above is the first report on resolving the polarization orientations of the mixed phases in highly-strained BFO thin films using AR-PFM. Since the electric and magnetic order parameters are intimately coupled, the exact polarization orientations in MPNs should shed light on the magnetoelectric coupling effect in this prototypical multiferroic. Furthermore, guided by the results from AR-PFM, we have demonstrated decisive manipulation of the MPNs formation by controlling the motion of a biased probe, which paves the way for future study on the electrical-controllable functionalities of the strain-driven MPB in BFO. The IP PFM study on the electrical switching process also helps us to gain more insight into the field-induced phase transitions between competing phases in BFO thin films.

**Materials and Methods.** High-strained BFO films with thicknesses of 20-80 nm was deposited on (001)-oriented LAO substrates using pulsed laser deposition. The growth temperature was fixed at 650 °C with oxygen pressure around 100 mTorr. The growth rate was about 1.5 nm/min at a laser repetition rate of 5 Hz. The phase purity of the films was confirmed by X-ray diffractometer (Shimadzu XRD-6000). The structural analyses using high-resolution RSMs were carried out at Singapore Synchrotron Light Source ($\lambda$ = 1.5405 Å). The RSMs were plotted in reciprocal lattice units (r.l.u.) of the LAO substrate (1 r.l.u. = $2\pi/3.789$ Å$^{-1}$). AR-PFM was performed on a commercial atomic force microscope system (Asylum Research MFP-3D) using DPE probe (Mikromash) with a spring constant of ~5 N/m. During PFM imaging, ac voltage with amplitude of 2 V and frequency of 10 kHz was applied to the probe. The electrical switching was carried out by applying -9 V dc bias to the probe at a scan rate of ~1 μm/s. The sample rotation angle in AR-PFM was calibrated using the <100> twinning boundaries arising from LAO substrate and the <110> domain walls in parent $M_C$ phase as well.

ACKNOWLEDGMENT

The authors acknowledge the support from Nanyang Technological University and Ministry of Education of Singapore under project number ARC 16/08. Partial support from National Research Foundation of Singapore under project NRF-CRP5-2009-04 is also acknowledged. Prof. G. L. Yuan acknowledges support from National Natural Science Foundation of China under project No. 11134004.

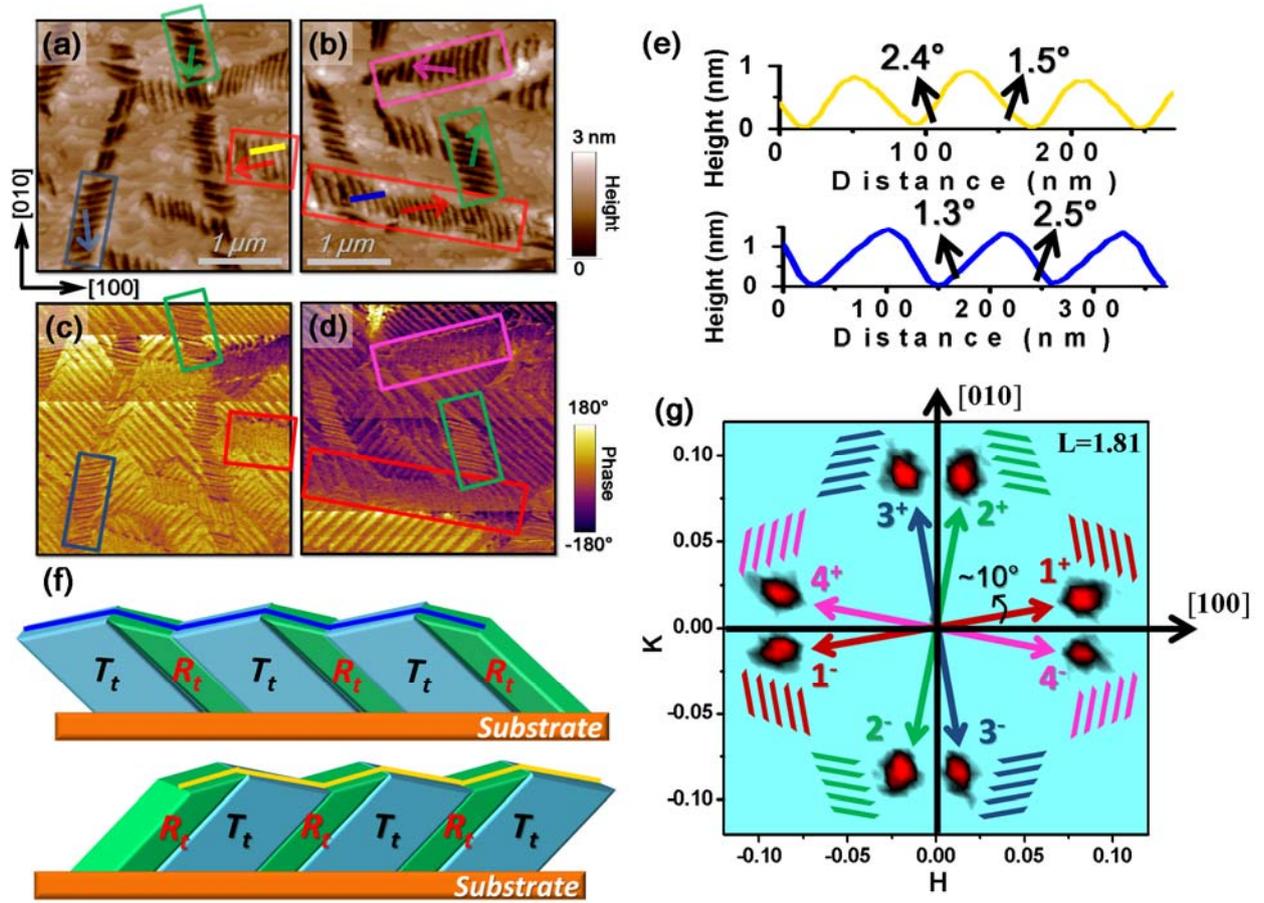

**Figure 1.** Topographic images (a)&(b) and corresponding IP PFM images (c)&(d) in highly-strained BFO thin film with MPNs. The arrows and boxes in (a)&(b) denote respective MPNs as shown in (d). (e) Corresponding height profiles along the solid lines shown in (a). (f) Schematics of the phase mixtures shown in (e). (g) Cross-sectional H-K map with L fixed at the OP lattice constant of $R_{tilt}$ phase. The eight peaks correspond to the eight structural vectors of stripe-like MPNs in highly-strained BFO thin films.

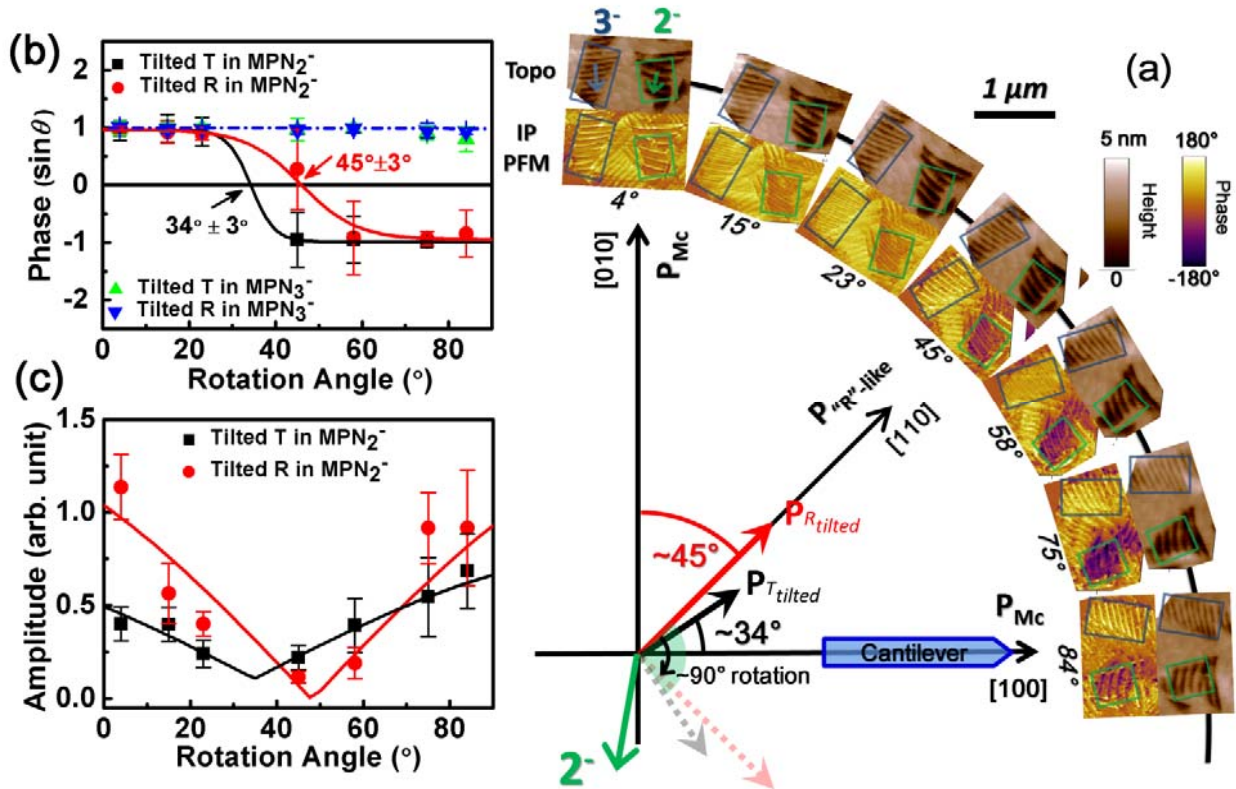

**Figure 2.** AR-PFM study of two specific MPN variants: $MPN_2^-$ and $MPN_3^-$. (a) Topographic and IP PFM images scanned at different sample rotation angles with regard to the cantilever. The coordinate at bottom left illustrates the IP polarization orientations of $T_{tilted}$ and $R_{tilted}$ phases in $MPN_2^-$ and their relationship with the parent $M_C$ phase. (b) Angular-dependent PFM phase signals for both $T_{tilted}$ and $R_{tilted}$ phases in $MPN_2^-$ and $MPN_3^-$. (c) Normalized angular-dependent amplitude signals for $T_{tilted}$ and $R_{tilted}$ phases in $MPN_2^-$.

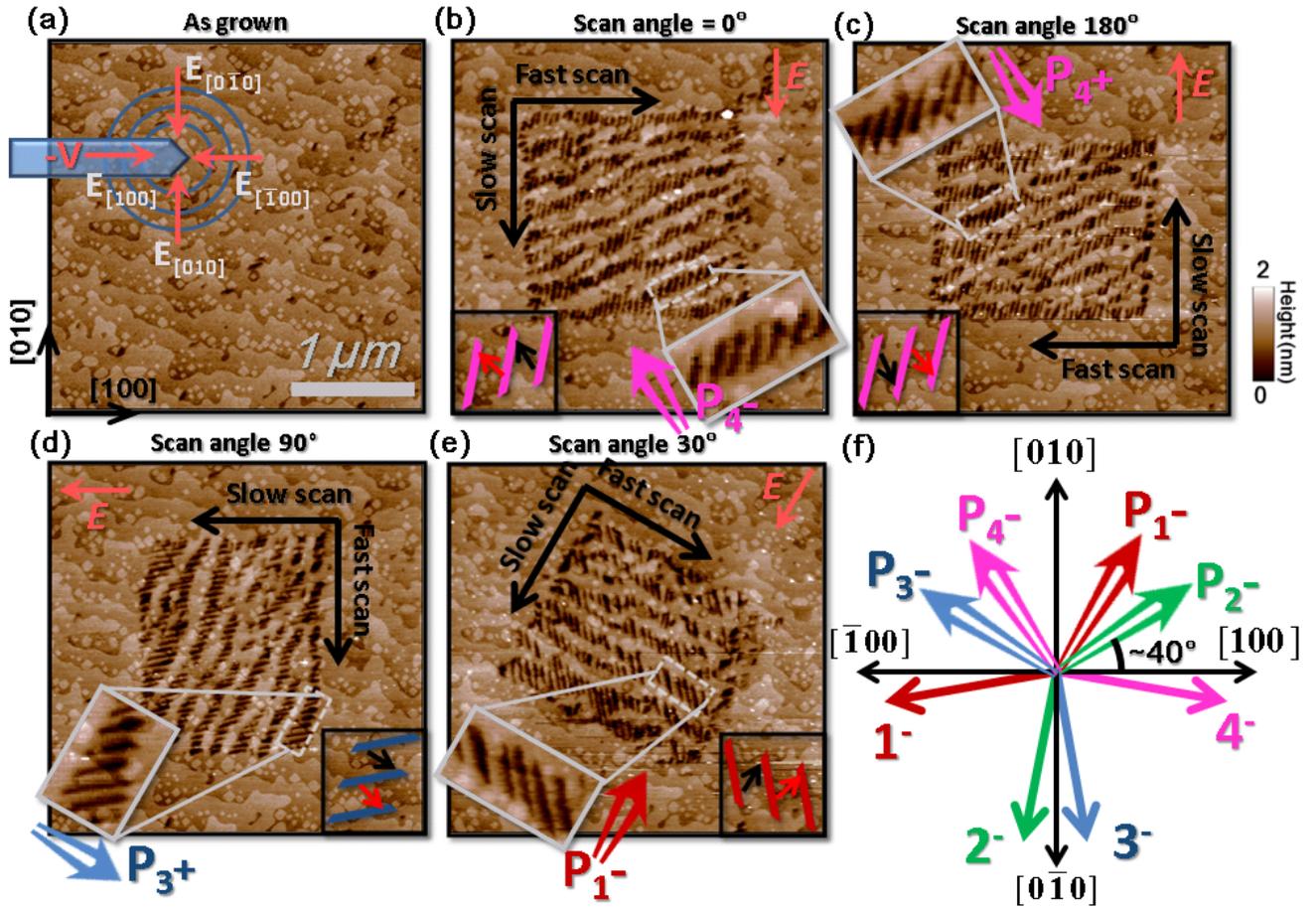

**Figure 3.** Topographic images of (a) as-grown BFO film and after electrical writing with scan angle of (b) 0°, (c) 180°, (d) 90° and (e) 30°, respectively. The inset of (a) shows the IP components of electric field induced by a negatively-biased probe. The insets of (b)-(e) display enlarged features of the field-induced MPNs, together with the average polarization directions. The detail polarization configurations for each MPN are schematically shown. The fast and slow scan directions and the IP electric field at the rear side of the slow scan direction are indicated in (b)-(e) as well. (f) IP polarization map denoting structural vectors of four negative MPN variants and their corresponding average polarization orientations of the $T_{tilted}$ and $R_{tilted}$ phases.

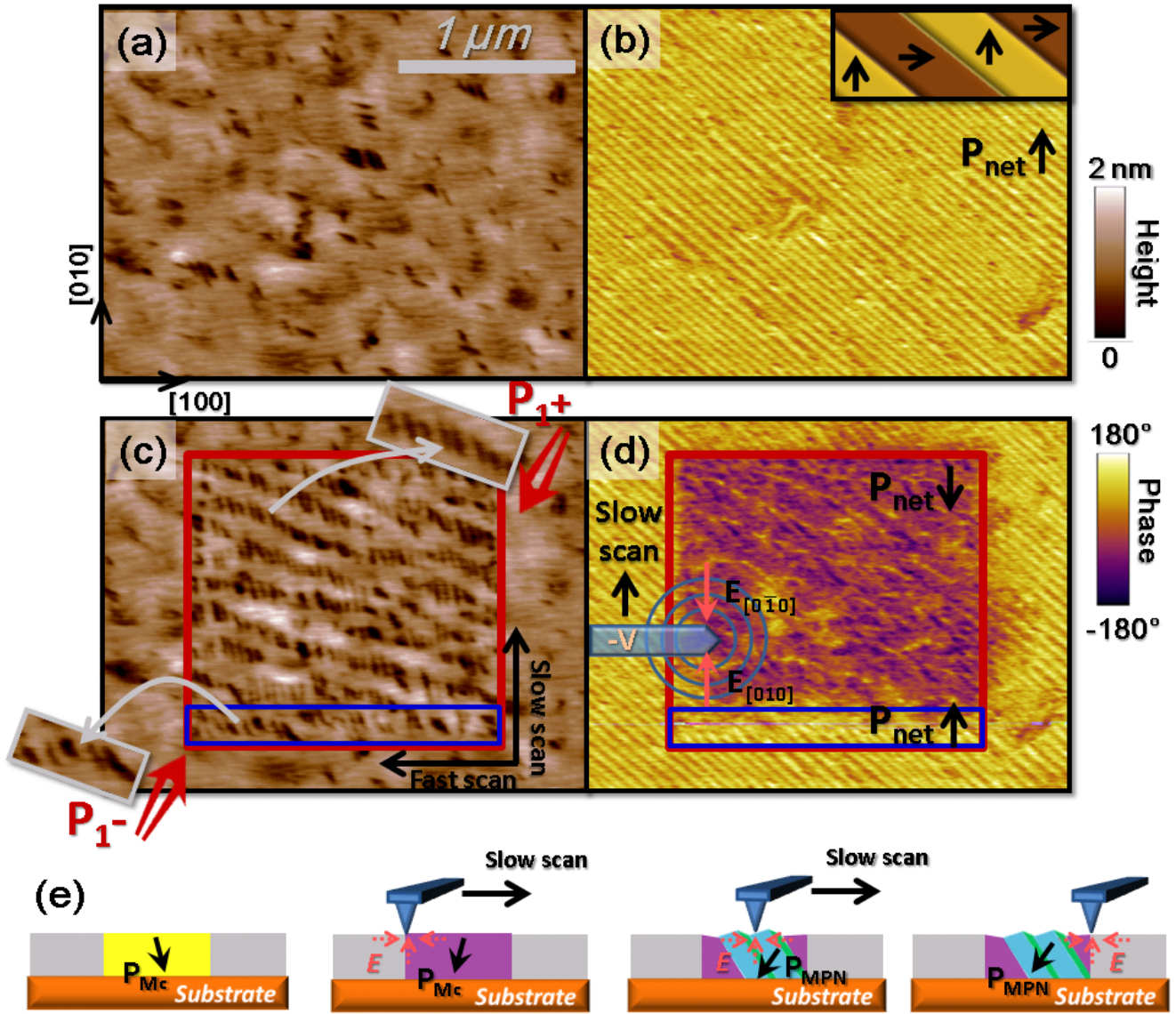

**Figure 4.** (a) Topographic and (b) IP PFM images of as-grown BFO film grown on LAO. (c) Topographic and (d) IP PFM image after -9 V electrical writing within the square box. The inset of (b) illustrates the polarization configuration of the $M_C$ phase with stripe domain pattern. Enlarged panels are shown in (c) for two type of field-induced MPNs. The probe motion and IP electric field are indicated in (d) as well. (e) Schematic showing the scenario of electric-field-induced phase transition between $M_C$ phase and MPNs during electrical writing process.

**SUPPORTING INFORMATION**

# Characterization and Manipulation of Mixed Phase Nanodomains in Highly-strained BiFeO$_3$ Thin Films

*Lu You,[†] Zuhuang Chen,[†] Xi Zou,[†] Hui Ding,[†] Weigang Chen,[†] Lang Chen,[†] Guoliang Yuan,[‡]\* and Junling Wang[†]\**

[†] School of Materials Science and Engineering, Nanyang Technological University, Singapore, 639798

[‡] Department of Materials Science and Engineering, Nanjing University of Science and Technology, Nanjing, 210094, China

\*E-mail: yuanguoliang@mail.njust.edu.cn, jlwang@ntu.edu.sg

**Figure S1**

A typical (002) H-L RSM of mixed-phase BFO film is shown in Figure S1(a). The diffraction peaks of different phases can be roughly divided into two L levels. The first one located at L = 1.62 (c = 4.67 Å), including the untilted $M_C$ matrix phase and the $T_{tilt}$ phase in the MPNs. The second one is $R_{tilt}$ with the L = 1.81 (c = 4.18 Å). By fixing the L at these two levels, the cross-sectional H-K maps can be obtained, revealing the IP information of the structural variants. As shown in Figure S1(b)&(c), both of the H-K maps of the $T_{tilt}$ and $R_{tilt}$ phases exhibit eight peaks, indicating eightfold degeneracy of the IP structural variants. The OP tilting and IP rotation angles of the $T_{tilt}$ and $R_{tilt}$ phases are in good agreement with the results derived from surface morphologies. It is worth noting that for each MPN the tilting angles of the $T_{tilt}$ and $R_{tilt}$ phases are actually in opposite directions with regard to the film surface normal as shown in Figure 1(e). As a result, the $R_{tilt}$ variants (e.g. $1^+$) always pair with the $T_{tilt}$ variants with opposite superscripts (e.g. $1^-$) to form MPNs. This can be justified by the fact that their peak intensities are always consistent with each other. For example, the $3^-$ and $4^-$ variants in $R_{tilt}$ phase have relative low intensities. Therefore, the $3^+$ and $4^+$ variants in $T_{tilt}$ are also weak. In the context, for simplification we refer the MPN variants using the label in the $R_{tilt}$ phase.

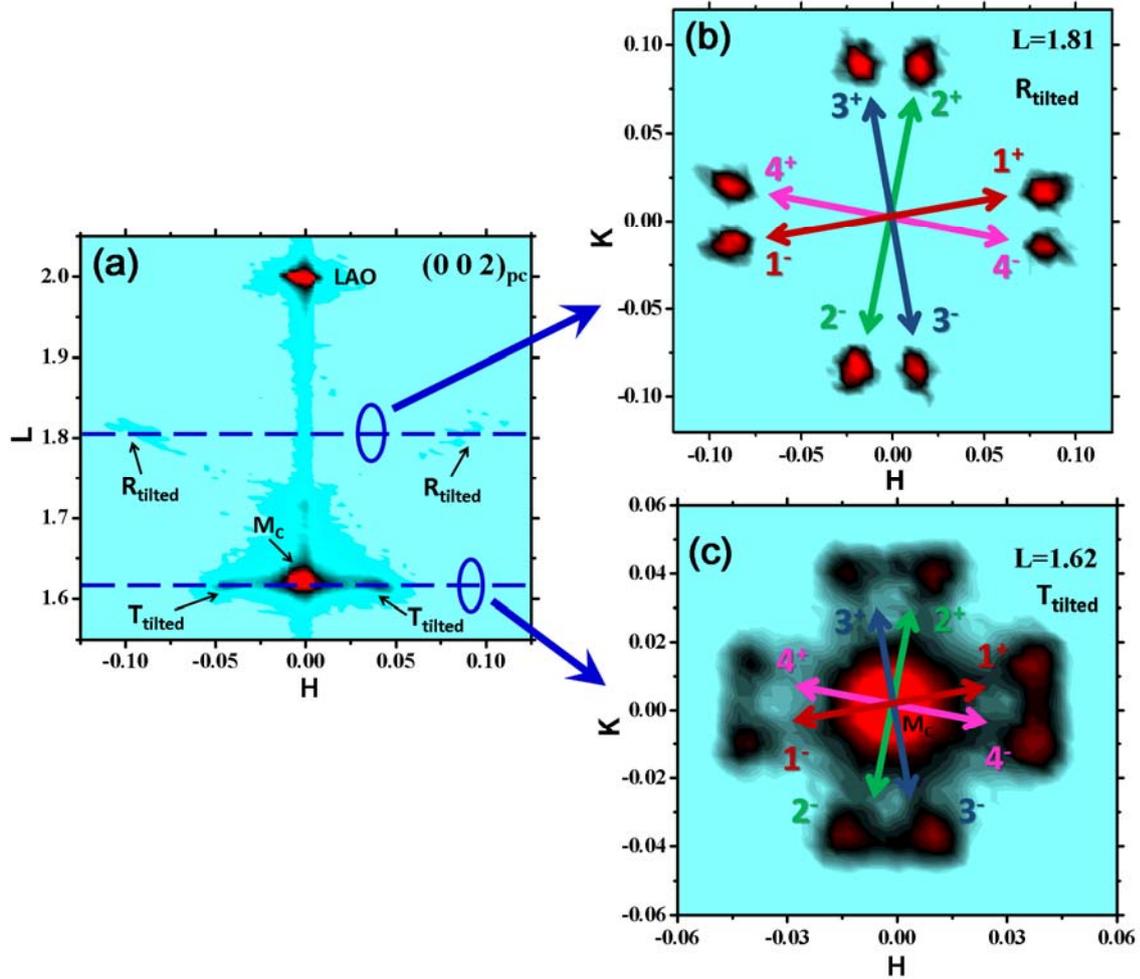

**Figure S1.** (a) (002) H-L RSM of 60 nm-thick BFO thin film grown on LAO. (b) Cross-sectional H-K map with L fixed at the OP lattice constant of $R_{tilt}$ phase. (c) Cross-sectional H-K map with L fixed at the OP lattice constant of $T_{tilt}$ phase.

**Figure S2**

The OP PFM images show no angular dependence. The phase image displays a uniform purple tone, indicating that all the OP polarizations are pointing downward. Enhanced piezoelectric response can be found at the mixed phase regions. As shown in Figure S2(g), the maxima of the piezoelectric response actually locate at the boundaries between the $T_{tilted}$ and $R_{tilted}$ phases, in agreement with previous reports.[S1, S2]

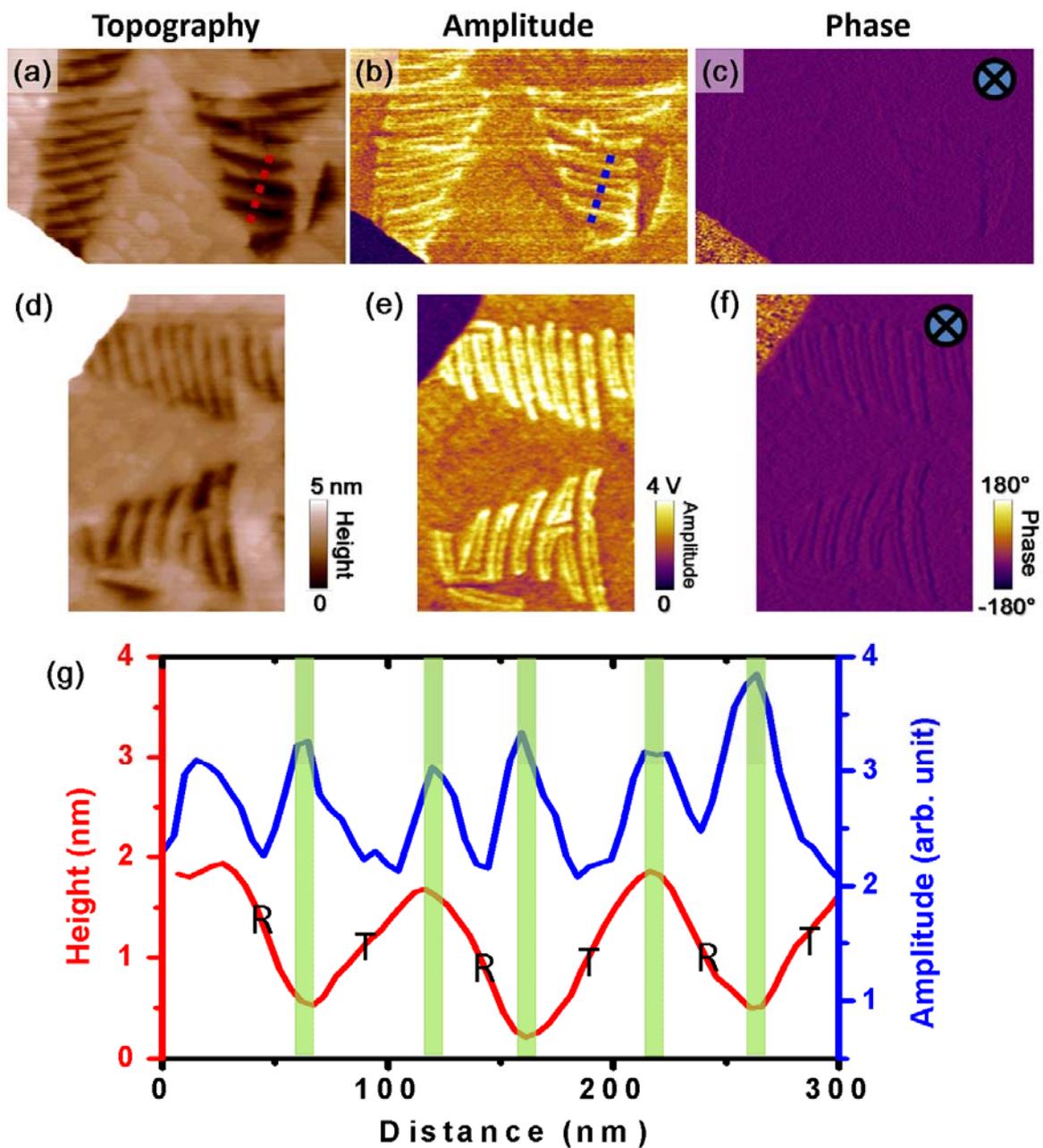

**Figure S2.** (a) & (d) Topography, (b) & (e) OP PFM amplitude, (c) & (f) OP PFM phase images scanned at the sample rotation angle of around 0º and 90º, respectively. (g) Respective line profiles denoted in (a) and (b).

**Figure S3**

To corroborate the conclusions obtained in the AR-PFM study, we have performed similar tests on other samples as illustrated in Figure S3. The angular-dependent phase evolutions can be monitored for four types of MPNs as denoted by the color boxes. We still focused on the $MPN_2^-$ as it is the only MPN that changes phase contrast within the rotation range. The PFM phase signal versus rotation angle can be plotted for two sets of $MPN_2^-$, as indicated by large (region1) and small (region2) boxes in Figure S3(a). The phase signals changed sign at ~33° for $T_{tilt}$ phase and at ~42° for $R_{tilt}$ phase, in good agreement with previous results. According to the polarization map drawn in Figure S3(f), we can examine the rest of MPNs, all of which didn't change phase contrast in the PFM images throughout the rotation range.

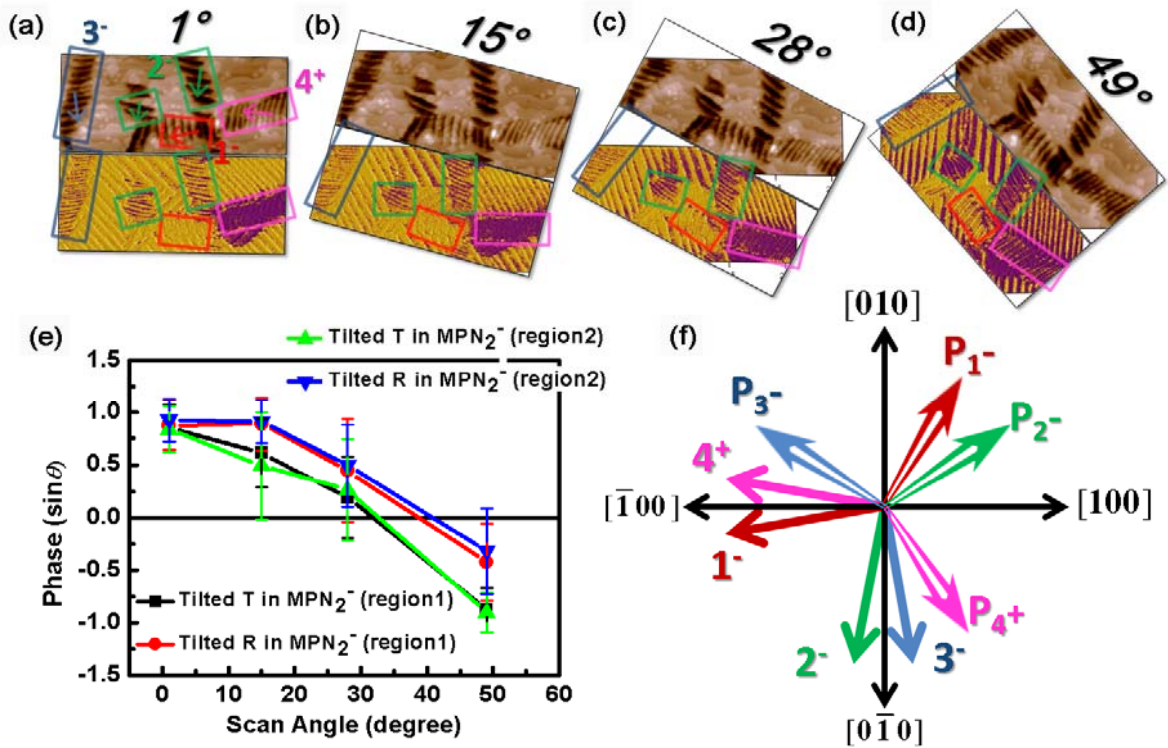

**Figure S3.** (a)-(d) The topographic and AR-PFM images of BFO thin film with MPNs. The arrows and boxes denote corresponding MPN variants as shown in (f). (e) Angular-dependent PFM phase

signals for both $T_{tilted}$ and $R_{tilted}$ phases in two sets of $MPN_2^-$. (f) IP polarization map denoting structural vectors of related MPN variants shown in (a) and their corresponding average polarization orientations of the $T_{tilted}$ and $R_{tilted}$ phases.

**Figure S4**

Since the OP polarization in the as-grown BFO film is pointing downward, a positive bias applied to the probe will transform the tilted phase mixture into parent $M_C$ phase. As shown in Figure S4a) & (c), the minor mixed phases featured by the depressions on the surface have been restore within the writing area. Concomitantly, the ferroelectric domain structure of the major $M_C$ phase has been modified by the IP electric field through ferroelastic switching. The net polarization in the switched area follows the electric field at the rear side of the slow scan direction. And the switching is mainly accomplished by IP 90° rotation of the polarization in each stripe domain. This result is very similar to those demonstrated in conventional "R"-like BFO thin films.[S3, S4]

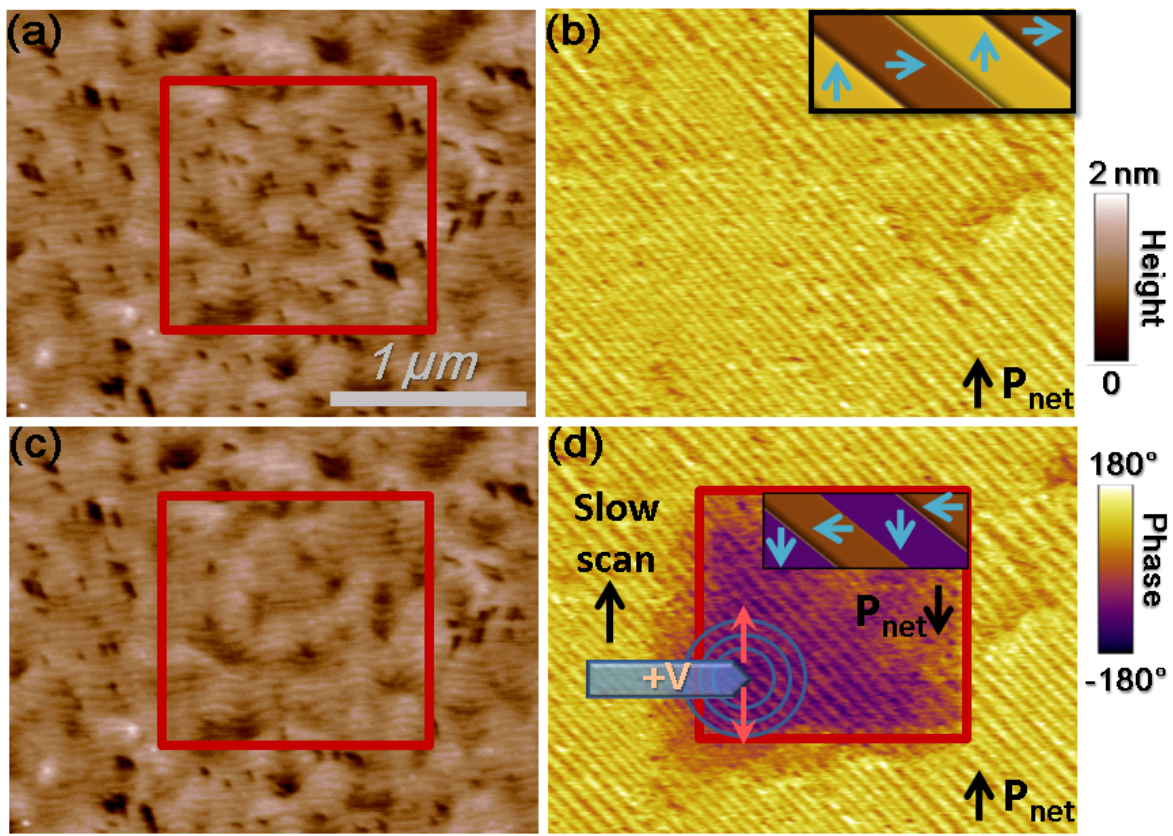

**Figure S4.** (a) Topography and (b) IP PFM image of as-grown BFO film on LAO. (c) Topography and (d) IP PFM image after +9 V electrical switching. The insets of (b) & (d) illustrate the polarization

configurations of the stripe domain patterns. The probe motion and IP electric field are indicated in (d) as well.